\documentclass{article}
\usepackage{amssymb,amsmath,cite,bm}
\def\rd{\mathrm{d}}
\begin{document}
\begin{titlepage}
\begin{center}
{\large \textbf{Entropy as a measure of diffusion}}

\vspace{2\baselineskip}
{\sffamily Amir~Aghamohammadi\footnote{e-mail: mohamadi@alzahra.ac.ir},
Amir~H.~Fatollahi\footnote{e-mail: fath@alzahra.ac.ir;~~ tel/fax: +98-21-88613937},
Mohammad~Khorrami\footnote{e-mail: mamwad@mailaps.org}
Ahmad~Shariati\footnote{e-mail: shariati@mailaps.org}
}

\vspace{2\baselineskip}
{\it Department of Physics, Alzahra University, Tehran 19938-93973, Iran}

\end{center}

\begin{abstract}
\noindent The time variation of entropy, as an alternative to
the variance, is proposed as a measure of the diffusion rate.
It is shown that for linear and time-translationally invariant systems
having a large-time limit for the density, at large times
the entropy tends exponentially to a constant. For systems with
no stationary density, at large times the entropy is logarithmic with
a coefficient specifying the speed of the diffusion. As an example,
the large time behaviors of the entropy and the variance are
compared for various types of fractional-derivative diffusions.
\end{abstract}

\vspace{2\baselineskip}



\end{titlepage}
\newpage
\section{Introduction}
Diffusion-like processes could be roughly characterized as those
processes in which some kind of density evolves so that its {\em width}
increases with time. But to make this more rigorous, one needs to specify
what exactly is meant by {\em width}. One choice is the variance, defined as
\begin{equation}\label{08.1}
\mathrm{Var}(X):=\langle X^2\rangle-\langle X\rangle^2.
\end{equation}
It is well-known that in ordinary diffusion processes on vector spaces
(which are of course unbounded) the variance varies linearly with time.
One way to compare other diffusion-like processes is then to calculate
the time dependence of the variance, to see whether its large time behavior
grows faster or slower than linear time dependence. A group of
diffusion-like processes are those called anomalous diffusion, which result
in a power-law for the large time dependence of the variance.  Investigations
of such processes include, for example,
\cite{meka2000,hughes,bouch,meka2004,vlahos,blu1986,shles}.
There are, however, diffusion-like processes for which
the variance is constant or blows up, an example of which is
the diffusion-like process in which the time derivative in
the ordinary diffusion is substituted with a Weyl fractional
derivative. There are also systems for which ordinary
variance is not defined. An example is diffusion on a compact
space, like a circle or sphere. In a recent article  it is shown that
for linear diffusion equations which preserve both time- and space-translational
invariance,  all connected moments are at most linear functions of time \cite{MSAF}.

An alternative for the variance as a means of quantifying
the diffusion, is the entropy. Unlike the variance, entropy
is defined for any density, and the way it grows could be
a measure of how fast the diffusion occurs. Here by how fast,
it is meant apart from a coefficient which could be absorbed in
the definition of time. In \cite{scagri}, entropy has been
used to detect scaling properties in the evolution of
a one dimensional system.

The aim of this paper is to present the time variation of
entropy as an alternative to the time derivative of moments of
probability for investigations of diffusion, specially in cases when
the moments are not well-defined, as it could happen in some forms
of anomalous diffusion.

The scheme of the paper is the following. In section 2 the entropy
corresponding to a density is defined. Section 3 is on the
time variation of entropy for evolutions corresponding to which
there is a stationary density. It is shown there that for
linear and time-translationally invariant systems, for large
times the entropy tends exponentially to a constant value.
Section 4 is on the time variation of entropy for evolutions
corresponding to which there is no stationary density. It is
shown that for a group of such processes, at large times
the entropy behaves like the logarithm of the time, with
a coefficient specifying the speed of the diffusion.
Some anomalous diffusion processes are investigated using
the entropy instead of the variance. It is shown that
different types of fractional derivatives with the same
index, result in similar large time behaviors for the entropy,
while they could result in completely different behavior for
the variance. Finally, section 5 is devoted to the concluding remarks.
\section{The entropy}
Consider a state space the points of which are denoted
by $\bm{r}$. The entropy corresponding to the dimensionless density
$\rho$ is denoted by $S$, and defined as
\begin{equation}\label{08.2}
S(\rho):=-\int\rd V\;\rho(\bm{r})\,\ln[\rho(\bm{r})],
\end{equation}
where integration is over the state space. The state space could be
continuous or discrete, in the latter case integration is substituted
with a summation. One has
\begin{equation}\label{08.3}
\int\rd V\;\rho(\bm{r})=1.
\end{equation}
Corresponding to any density $\rho_0$ (including the stationary
density, if such a thing exists), one defines $S_{\rho_0}$ as
\begin{equation}\label{08.4}
S_{\rho_0}(\rho):=-\int\rd V\;\rho(\bm{r})\,\ln\left[\frac{\rho(\bm{r})}{\rho_0(\bm{r})}\right].
\end{equation}
The motivation for such a definition is the following. The Shannon entropy for
a system of discrete states is
\begin{equation}\label{08.5}
S_1=-\sum_i p_i\,\ln p_i,
\end{equation}
where $p_i$ is the probability of the state $i$. As $p_i$'s
are all nonnegative and less than or equal to one, the above
entropy is positive. To write something similar for a system
of continuous states, one can divide the space into cells of
volume $\Delta V_i$, and define the entropy like
\begin{equation}\label{08.6}
S_2=-\sum_i(\rho_i\,\Delta V_i)\,\ln(\rho_i\,\Delta V_i),
\end{equation}
where $\rho_i$ is the probability density in the $i$'th cell.
One would expect to get an exact result, in the limit the cell
sizes tend to zero. However, $S_2$ tends to (positive) infinity,
as the cell sizes tend to zero. Moreover, it does depend on
the way the space is divided into cells. One could say that the space
can be divided into cells of {\em equal} volume $v$, and a term
$(-\ln v)$ be subtracted from $S_2$, before tending $v$ to $0$,
in order to get a finite result for the entropy. The entropy
obtained this way, however, is not necessarily positive, as it is
the Shannon entropy from which a positive infinite term is subtracted
to make it regular. Another point is that the volume of cells
does depend on the volume element (or the choice of coordinates).
So the choice of cells of equal volume is not unambiguous. One way
to define (remove) this unambiguity, is to use a reference density.
Suppose the cells are selected so that the probability of each cell
with the reference density $\rho_0$ is constant:
\begin{equation}\label{08.7}
(\rho_0)_i\,\Delta V_i=p.
\end{equation}
One would then have
\begin{equation}\label{08.8}
S_2=-\sum_i(\rho_i\,\Delta V_i)\,\ln\frac{p\,\rho_i}{(\rho_0)_i}.
\end{equation}
Sending the cell sizes to zero (but maintaining the condition
that the probability of cells with regard to $\rho_0$ be equal)
is then equivalent to sending $p$ to zero. To regularize $S_2$,
in the limit $p$ tending to zero, one subtracts a constant term
$(-\ln p)$ from it:
\begin{equation}\label{08.9}
S'_2=\ln p-\sum_i(\rho_i\,\Delta V_i)\,\ln\frac{p\,\rho_i}{(\rho_0)_i}.
\end{equation}
The above, in the limit of cell sizes tending to zero, is (\ref{08.4}).
This entropy differs from $S_2$ by a constant (infinite), and
is no longer positive definite.

As the function $\mathcal{S}$ with
\begin{equation}\label{08.10}
\mathcal{S}(\xi):=-\xi\,\ln\xi
\end{equation}
is concave, $S_{\rho_0}(\rho)$ is negative, unless $\rho$ is the same as
$\rho_0$. For $\rho$ slightly different from $\rho_0$, one has
\begin{align}\label{08.11}
S_{\rho_0}(\rho)&=-\int\rd V\;[\rho_0+(\rho-\rho_0)]\,\ln\left(1+\frac{\rho-\rho_0}{\rho_0}\right)
,\nonumber\\
&=-\int\rd V\;[\rho_0+(\rho-\rho_0)]\,\left[\frac{\rho-\rho_0}{\rho_0}
-\frac{1}{2}\,\left(\frac{\rho-\rho_0}{\rho_0}\right)^2\right]+\cdots,\nonumber\\
&=-\int\rd V\;\left[(\rho-\rho_0)+\frac{1}{2}\,\frac{(\rho-\rho_0)^2}{\rho_0}\right]+\cdots.
\end{align}
The first order term is zero, as (\ref{08.3}) holds for both $\rho$ and $\rho_0$. So,
\begin{equation}\label{08.12}
S_{\rho_0}(\rho)=-\frac{1}{2}\,\int\rd V\;\frac{[\rho(\bm{r})-\rho_0(\bm{r})]^2}{\rho_0(\bm{r})}.
\end{equation}

The evolution of entropy with time ($t$) is, of course, obtained through
the time evolution of the density. There are time evolutions for densities,
which do have a stationary density; and there are time evolutions for
densities, which do not have stationary densities. An example of
the first is ordinary diffusion on a compact space. An example of
the second is ordinary diffusion on a non-compact space. The large time
behavior of the entropy corresponding to these two kinds of evolution is
different.
\section{Time variation if there is a stationary density}
If there is a stationary density $\rho_0$, then as mentioned above	,
$S_{\rho_0}$ has a maximum which is attained if the density is
the same as the stationary density. For large times, the density
$\rho$ approaches $\rho_0$, so that one could use (\ref{08.12})
for the entropy:
\begin{equation}\label{08.13}
S(t)=-\frac{1}{2}\,\int\rd V\;\frac{[\rho(t,\bm{r})-\rho_0(\bm{r})]^2}{\rho_0(\bm{r})}.
\end{equation}
If the evolution of the density is linear and time-translationally invariant, then
\begin{equation}\label{08.14}
\dot\rho(t,\bm{r})=[H(\bm{r},\bm{D})]\,\rho(t,\bm{r}),
\end{equation}
where $\bm{D}$ is differentiation with respect to $\bm{r}$, meaning that
$[H(\bm{r},\bm{D})]$ is a differential operator. An example is
the simple diffusion, for which $H$ is the Laplacian, or $(\bm{D}\cdot\bm{D})$.

For the stationary density $\rho_0$,
\begin{equation}\label{08.15}
[H(\bm{r},\bm{D})]\,\rho_0(\bm{r})=0.
\end{equation}
If the evolution is so that the stationary density is the large-time density,
that is, if all initial densities tend to $\rho_0$ at large times, and if
all of the eigenvalues of $H$, apart from zero, have real values smaller
than a negative value, then
\begin{equation}\label{08.16}
\rho(t,\bm{r})=\rho_0(\bm{r})+\mathrm{Re}[\rho_1(\bm{r})\,\exp(E_1\,t)]+\cdots,
\qquad\mbox{for large times,}
\end{equation}
where $E_1$ is that eigenvalue of $H$ which has the largest negative real part,
and $\rho_1$ is the eigenfunction of $H$ corresponding to $E_1$. One then arrives at
\begin{equation}\label{08.17}
S(t)=-\exp[2\,\mathrm{Re}(E_1)\,t]\,s_1(t),\qquad\mbox{for large times,}
\end{equation}
where $s_1(t)$ is a positive constant if $E_1$ is real, and an oscillatory
nonnegative non-decaying function if $E_1$ is not real.
\subsection{Example: diffusion on compact spaces}
The evolution corresponding to ordinary diffusion is
\begin{equation}\label{08.18}
\dot\rho=\bm{D}\cdot\bm{D}\,\rho.
\end{equation}
If the state space is compact and boundary-less, the Laplacian operator
$\bm{D}\cdot\bm{D}$ has a zero eigenvalue corresponding to
the eigenfunction $\rho_0$, which is a constant function.
One can also normalize this constant so that its integral is one.
Moreover, the other eigenvalues of Laplacian are negative, so
the density approaches $\rho_0$ at large times. The behavior of
the entropy at large times is then determined by the largest
negative eigenvalue of the Laplacian. Examples of such compact spaces
are the circle and the (two-dimensional) sphere. One has
\begin{equation}\label{08.19}
E_1=\begin{cases}\displaystyle{-\frac{1}{a^2}},&\mbox{circle of radius $a$}\\ \\
\displaystyle{-\frac{2}{a^2}},&\mbox{sphere of radius $a$}
\end{cases}.
\end{equation}
\section{Time variation if there is no stationary density}
Consider the equation
\begin{equation}\label{08.20}
\dot\rho(t,\bm{r})=[H(t,\bm{r},\bm{D})]\,\rho(t,\bm{r}).
\end{equation}
There are cases for which there is some function $\bm{f}$ so that the solution to
the above equation has the property that
\begin{equation}\label{08.21}
\det\left[\frac{\partial\bm{f}(t,t_0,\bm{r})}{\partial\bm{r}}\right]\,\rho[t,\bm{f}(t,t_0,\bm{r})]
=\rho(t_0,\bm{r}).
\end{equation}
This means the points around $\bm{r}$ at the time $t_0$, go to the
region around the point $\bm{f}(t,t_0,\bm{r})$ at the time $t$. The determinant
at the left-hand side takes into account the fact that the volume of the region
is changing.

For such cases,
\begin{align}\label{08.22}
S(t)&=-\int\rd V\;\rho(t,\bm{r})\,\ln[\rho(t,\bm{r})],\nonumber\\
&=-\int\rd V'\;\det\left[\frac{\partial\bm{f}(t,t_0,\bm{r}')}{\partial\bm{r}'}\right]
\,\rho[t,\bm{f}(t,t_0,\bm{r}')]\,\ln\{\rho[t,\bm{f}(t,t_0,\bm{r}')]\},\nonumber\\
&=-\int\rd V'\;\rho(t_0,\bm{r}')\,\ln[\rho(t_0,\bm{r}')]+\int\rd V'\;\rho(t_0,\bm{r}')\,
\ln\left\{\det\left[\frac{\partial\bm{f}(t,t_0,\bm{r}')}{\partial\bm{r}'}\right]\right\},\nonumber\\
&=S(t_0)+\int\rd V\;\rho(t_0,\bm{r})\,
\ln\left\{\det\left[\frac{\partial\bm{f}(t,t_0,\bm{r})}{\partial\bm{r}}\right]\right\},
\end{align}
or
\begin{equation}\label{08.23}
\dot S=\int\rd V\;\rho(t_0,\bm{r})\,\bm{D}\cdot\bm{u}(t,\bm{r}),
\end{equation}
where $\bm{f}$ is the flux corresponding to the vector field $\bm{u}$:
\begin{equation}\label{08.24}
\frac{\partial\bm{f}(t,t_0,\bm{r})}{\partial t}\Big|_{t=t_0}=\bm{u}(t,\bm{r}).
\end{equation}
But (\ref{08.21}) is equivalent to
\begin{equation}\label{08.25}
\dot\rho(t,\bm{r})=-\bm{D}\cdot[\bm{u}(t,\bm{r})\,\rho(t,\bm{r})].
\end{equation}
This corresponds to a deterministic evolution of $\bm{r}$,
through the vector field $\bm{u}$, and does not involve any diffusion.
However, it could happen that for large times, when the density becomes
slowly-varying with $\bm{r}$, equation (\ref{08.21}) holds approximately.
Consider the following transformations:
\begin{align}\label{08.26}
t&\to[\exp(\alpha)]\,t,\nonumber\\
\bm{r}&\to[\exp(\alpha\,R)]\,\bm{r},
\end{align}
where $\alpha$ is a parameter and $R$ is a constant matrix. If
(\ref{08.20}) is invariant under this transformation (for large $\alpha$),
then one could take $\bm{f}$ to be
\begin{equation}\label{08.27}
\bm{f}\{[\exp(\alpha)]\,t_0,t_0,\bm{r}\}=\exp(\alpha\,R)\,\bm{r},
\end{equation}
so that
\begin{equation}\label{08.28}
\frac{\partial\bm{f}(t,t_0,\bm{r})}{\partial\bm{r}}=\exp\left(R\,\ln\frac{t}{t_0}\right),
\end{equation}
resulting in
\begin{equation}\label{08.29}
S(t)=S(t_0)+[\mathrm{tr}(R)]\,\ln\frac{t}{t_0},
\end{equation}
or
\begin{equation}\label{08.30}
\dot S=\frac{\mathrm{tr}(R)}{t}.
\end{equation}
Of course these hold for only large times.
\subsection{Example: anomalous diffusion with no stationary state}
Consider an evolution equation of the form
\begin{equation}\label{08.31}
D_{0\,(\beta_0)}\,\rho=[f(\bm{D})]\,\rho,
\end{equation}
where $D_{0\,(\beta_0)}$ is a fractional (time) derivative of
order $\beta_0$, while $\bm{D}$ is the space derivative.
$[f(\bm{D})]$ itself, could be fractional. Depending on
what kind of fractional derivative is used, $D_{0\,(\beta_0)}$
could depend on a time. That is, it could be that
$D_{0\,(\beta_0)}$ is not time invariant. But even if that is
the case, for large times that special time is unimportant, and one has
\begin{equation}\label{08.32}
D_{0\,(\beta_0)}\to[\exp(-\alpha\,\beta_0)]\,D_{0\,(\beta_0)},\qquad\mbox{for }t\to[\exp(\alpha)]\,t.
\end{equation}
One also has
\begin{equation}\label{08.33}
\bm{D}\to\bm{D}\,[\exp(-\alpha\,R)],\qquad\mbox{for }\bm{r}\to[\exp(\alpha\,R)]\,\bm{r}.
\end{equation}
So, if $R$ can be chosen so that
\begin{equation}\label{08.34}
f[\bm{D}\,\exp(-\alpha\,R)]=[\exp(-\alpha\,\beta_0)]\,f(\bm{D}),
\end{equation}
(for large values of $\alpha$), then (\ref{08.29}) and (\ref{08.30})
would hold. As an example, for
\begin{equation}\label{08.35}
D_{0\,(\beta_0)}\,\rho=\sum_ja_j\,D_{j\,(\beta_j)}\,\rho,
\end{equation}
where $D_{j\,(\beta_j)}$ is some (possibly fractional) differentiation of order
$\beta_j$ with respect to $r^j$, is is easily seen that the transformation
$R$ satisfying (\ref{08.34}) is the following.
\begin{equation}\label{08.36}
R\,\begin{pmatrix}r^1\\ \vdots\\ r^n\end{pmatrix}=
\begin{pmatrix}(\beta_0/\beta_1)\,r^1\\ \vdots\\(\beta_0/\beta_n)\,r^n\end{pmatrix},
\end{equation}
So that
\begin{equation}\label{08.37}
\dot S=\left(\sum_j\frac{1}{\beta_j}\right)\,\frac{\beta_0}{t}.
\end{equation}
Normal diffusion corresponds to $\beta_0=1$ and $\beta_j=2$, resulting in
the following large time behavior.
\begin{equation}\label{08.38}
\dot S=\frac{d}{2\,t},
\end{equation}
where $d$ is the dimension of space.

As special cases, consider the followings.
\subsubsection{Anomalous diffusion with fractional time derivative}
Consider an evolution of the form
\begin{equation}\label{08.39}
D_{0\,(\beta_0)}\,\rho=\bm{D}\cdot\bm{D}\,\rho,
\end{equation}
where $D_{0\,(\beta_0)}$ could be the Caputo or Weyl derivative of order
$\beta_0$; see \cite{frac1,frac2}. for example. For the Caputo derivative,
the variance varies as $t^\beta_0$:
\begin{equation}\label{08.40}
[\mathrm{Var}^\mathrm{C}(\bm{r})](t)=[\mathrm{Var}^\mathrm{C}(\bm{r})](0)+
\frac{2\,d\,t^{\beta_0}}{\Gamma(1+\beta_0)},
\end{equation}
where $d$ is the dimension of the space. For the Weyl derivative, the variance
is a finite constant for $\beta_0<1$; it is divergent at any positive time,
for $\beta_0>1$; and it varies linearly with time for $\beta_0=1$
(ordinary diffusion). For both kinds of derivatives, however, it is seen
from (\ref{08.37}) that for large times
\begin{equation}\label{08.41}
\dot S=\frac{\beta_0\,d}{2\,t}.
\end{equation}
\subsubsection{Anomalous diffusion with fractional space derivative}
L\'{e}vy flight is a stochastic, Markov process, which differ from
regular Brownian motion by the occurrence of extremely long jumps
the probability distribution of which is heavy-tailed.
L\'{e}vy flight in continuum limit can be mapped onto
the diffusion equation with fractional space derivative.
One class of fractional space derivatives is defined as
\begin{equation}\label{08.42}
[\mathcal{F}(D_{(\beta)}\rho)](k):=-|k|^\beta\,[\mathcal{F}(\rho)](k),
\end{equation}
where $\mathcal{F}$ denotes the Fourier transform,
(see \cite{CGKM}, for example). If the evolution equation
for $\rho$ (on a one-dimensional space) is
\begin{equation}\label{08.43}
D_0\rho=D_{(\beta)}\rho,
\end{equation}
then the entropy satisfies at large times
\begin{equation}\label{08.44}
\dot S=\frac{1}{\beta\,t}.
\end{equation}
As a special case, consider
\begin{equation}\label{08.45}
D_0\rho=D_{(1)}\rho,
\end{equation}
which corresponds to a special case of L\'{e}vy flights.
This results in
\begin{equation}\label{08.46}
\rho(t,x)=\int\rd y\;\frac{t\,\rho(0,y)}{\pi\,[t^2+(y-x)^2]}.
\end{equation}
For any positive $t$, the variance corresponding to the density
$\rho$ diverges. The entropy, however, does not. As both $\beta_0$ and
$\beta_1$ are equal to one, the large time behavior of the entropy is
\begin{equation}\label{08.47}
\dot S=\frac{1}{t}.
\end{equation}
\subsection{Example: anomalous diffusion with fractional space derivative, and a drift}
The evolution equation
\begin{equation}\label{08.48}
D_0\rho=D[(D U)\,\rho]+D_{(\beta)}\rho,
\end{equation}
corresponds to an anomalous diffusion with fractional space derivative,
combined with a drift, (\cite{CGKM}, for example). The drift is
the first term on the right hand side, so that the point $x$,
apart from a random motion, is subject to a drift governed by
\begin{equation}\label{08.49}
(D_0 x)(t)=-(D U)[x(t)].
\end{equation}
To investigate the large time behavior of the entropy for such systems,
the behavior of the potential $U$ for large values of its variable is needed.
Let us assume a power law:
\begin{equation}\label{08.50}
U(x)\sim-a\,|x|^c,\qquad\mbox{for large $|x|$}.
\end{equation}
Applying the transformations (\ref{08.26}) on (\ref{08.48}), one arrives at
\begin{equation}\label{08.51}
[\exp(-\alpha)]\,D_0\rho=\{\exp[(c-2)\,R\,\alpha]\}\,D[(D U)\,\rho]+
[\exp(-\beta\,R\,\alpha)]\,D_{(\beta)}\rho.
\end{equation}
Of course this should hold only for large (positive) $\alpha$ (and with
a positive $R$). So one obtains
\begin{equation}\label{08.52}
[\min(2-c,\beta)]\,R=1,
\end{equation}
where only a positive solution for $R$ is acceptable. If such a solution
for $R$ exists, then the large time behavior of the entropy is
\begin{equation}\label{08.53}
\dot S=\frac{R}{t}.
\end{equation}
It is assumed that $\beta$ is positive. The following cases occur.
\begin{itemize}
\item[\textbf{i}] $c<(2-\beta)$

It is the diffusion which determines the large time behavior of the system.
for large times,
\begin{equation}\label{08.54}
\dot S=\frac{1}{\beta\,t}.
\end{equation}
\item[\textbf{ii}] $(2-\beta)<c<2$, and $(a\,c)>0$

It is the potential which determines the large time behavior of the system.
The potential is repulsive, and for large times,
\begin{equation}\label{08.55}
\dot S=\frac{1}{(2-c)\,t}.
\end{equation}
\item[\textbf{iii}] $2<c$, and $(a\,c)>0$

The potential is repulsive, and the evolution blows up at a finite time.
\item[\textbf{iv}] $(2-\beta)<c$, and $(a\,c)<0$

It is the potential which determines the large time behavior of the system.
The potential is attractive, and for large times the system tends to a nonzero
density, so that the entropy does not grow indefinitely.
\end{itemize}
\section{Concluding remarks}
Entropy was introduced as a tool to study how fast the diffusion
in a system occurs. It was shown that there are systems for which
the evolution of variance is ill-defined, while the entropy
and its evolution are well-defined. The long time behavior of
the entropy was studied, and it was shown that there are
cases where the density tends to a stationary state
and the entropy tends exponentially towards its final value,
and there are cases where the entropy behaves like the logarithm
of time. Some anomalous diffusions were in particular
studied, to compare the information obtained from the entropy
to that obtained from the variance.
\\[\baselineskip]
\textbf{Acknowledgement}: This work was
supported by the Research Council of the Alzahra University.

\end{document}